\documentclass[12 pt]{article}

\usepackage{authblk}
\usepackage[style=bwl-FU,maxbibnames=9]{biblatex}
\usepackage{wrapfig}
\usepackage{caption}
\usepackage{hyperref}
\hypersetup{
    colorlinks=true,
    linkcolor=black,
    filecolor=magenta,  
    citecolor=black,
    urlcolor=blue,
    pdftitle={Overleaf Example},
    pdfpagemode=FullScreen,
    }

\usepackage[letterpaper,top=2cm,bottom=2cm,left=1.75cm,right=1.75cm,marginparwidth=1.75cm]{geometry}

\usepackage{graphicx}%
\usepackage{multirow}%
\usepackage{amsmath,amssymb,amsfonts}%
\usepackage{amsthm}%
\usepackage{mathrsfs}%
\usepackage[title]{appendix}%
\usepackage{xcolor}%
\usepackage{textcomp}%
\usepackage{manyfoot}%
\usepackage{booktabs}%
\usepackage{algorithm}%
\usepackage{algorithmicx}%
\usepackage{algpseudocode}%
\usepackage{listings}%

\usepackage{booktabs}
\usepackage{graphicx}
\usepackage{epstopdf}
\usepackage{upgreek}
\usepackage{multicol,multirow}
\usepackage{amsmath,amssymb,amsfonts}
\usepackage{mathrsfs}
\usepackage{amsthm}
\usepackage[figuresright]{rotating}
\usepackage{appendix}
\usepackage{ifpdf}
\usepackage[T1]{fontenc}
\usepackage{newtxtext}
\usepackage{newtxmath}
\usepackage{textcomp}
\usepackage{xcolor}
\usepackage{lineno}

\usepackage{amsmath,amsfonts,amssymb,mathtools}
\usepackage{gensymb}
\usepackage{siunitx}
\usepackage{cancel}

\title{Wind as Driver of Bird and Bat Abundance, Flight Direction, Altitude, and Speed on the North Atlantic Shelf}
\author[1,*]{Abigale Snortland}
\author[2]{Jeff Clerc}
\author[2]{Cris Hein}
\author[1]{Emma Cotter}

\affil[1]{\small{Pacific Northwest National Laboratory, Coastal Sciences Division, 1100 Dexter Ave N, Seattle, WA 98109}}
\affil[2]{\small{National Renewable Energy Laboratory, National Wind Technology Center, 19001 W. 119th Ave., Arvada, CO 80007}}
\affil[*]{\small{Corresponding Author, abigale.snortland@pnnl.gov}}

\begin{document}
\maketitle

\abstract{Quantifying the collision risk of birds and bats with offshore wind turbines requires an understanding of the drivers of flying animal behavior at offshore wind sites. This study presents a first methodological approach for analyzing avian radar and lidar data together, providing a framework for future analyses of offshore bird and bat movements that can be used to improve collision risk models. An omnidirectional S-band radar system was deployed on a research barge on the Northeastern Shelf of the United States ($40.9^\circ$ N, $70.79^\circ$ W) and collected data for a 5-week window during the 2024 autumn bird and bat migration. The barge also supported two profiling lidar systems that measured the wind speed and direction. Coupling the radar animal tracks with measured wind speed profiles revealed that wind is a driver of animal presence, flight direction, flight height, and flight speed. Further, a hierarchical clustering methodology was developed to investigate behavior by approximate animal size. For example, smaller animals had concentrated flight direction distributions aligned with the wind and flew at a variety of altitudes, whereas bigger animals flew in a wide variety of directions but were concentrated at low altitudes. Our results provide the first insights into animal behavior at offshore wind sites with paired radar and lidar data.     }




\maketitle

\section{Introduction}

Offshore wind farms are being developed in many areas of the world, including along the northeast coast of the United States. While offshore wind turbines provide a reliable source of renewable energy, they also have the potential to affect the ecosystems where they are deployed and the animals that inhabit those ecosystems \parencite{Fox2006}. One of the primary environmental risks associated with offshore wind development is the risk of flying animals (birds or bats) colliding with wind turbines. Exposure to offshore wind turbines and therefore collision risk is species-specific \parencite{Willmott2013,Kelsey2025}. Collision risk models are typically used to quantify collision risk for species or species groups. These models incorporate known parameters, including turbine dimensions and animal morphology (e.g., length, wingspan), but also require assumptions about animal behavior such as avoidance rate, which can be difficult to quantify \parencite{Band2012, Masden2016}. 

Several factors increase a species' vulnerability to collision risk, including daily and annual presence, flight activity metrics (e.g., flight height and flight speed; see \cite{Kelsey2025} for details on species collision vulnerability indices), and turbine operational state. The abundance of flying animals offshore varies with distance to shoreline, time of day, seasonal cycles (migration), and meteorological conditions \parencite{Drewitt2006, Davies2024, Fauchald2024}. Because flying animals are only at risk of collision if they are present in an offshore wind energy area, an understanding of how these factors influence animal abundance is needed to assess collision risk. These factors also influence animal behavior such as flight height and speed. Animals flying above or below the turbine rotor-swept zone (RSZ) are at lower risk for collision \parencite{Ainley2015,Bradaric2024}, and flight speed and direction influence collision probability and severity by affecting transit time through the rotor and the impact force \parencite{Holstrom2011,marques2014}. It is therefore critical to consider how animal behavior changes with meteorological conditions \parencite{
williams2024}. In addition to animal presence and behavior, collision risk will also vary depending on wind turbine operation, which changes with wind conditions. For example, blade rotation rates vary with wind speed, and turbines actively yaw to align with the prevailing wind direction. 

A variety of monitoring technologies and platforms can be used to investigate flying animal abundance and flight activity metrics offshore, including vessel-based visual surveys \parencite{Ainley2015, Johnston2015}, optical or thermal cameras \parencite{Hupop2006,RobinsonWillmot2023,Schneider2024}, telemetry \parencite{Ross-smith2016,gibb2018,Galtbalt2021,Masden2021,kumagai2023, Davies2024,Weiser2024,yordan2025}, acoustics \parencite{Peterson2014,Brabant2019,Lagerveld2023}, and radar \parencite{Hupop2006,Manola2020,VanErp2021,Leemans2022,Bradaric2024,Skov2026}. Vessel-based visual surveys can cover large areas and achieve species-level identification, but operations are limited to daylight periods with sufficient visibility and relatively low sea states suitable for marine operations. Further, visual surveys are limited in their ability to determine the flight height and speed of observed animals. Optical cameras can also achieve species-level identification under some conditions, and stereo-optical camera systems can determine flight height and flight speed, but their operation requires ambient light. Conversely, thermal cameras can operate without ambient light, but have more limited classification capabilities. Both types of cameras are limited in detection range and, when deployed near the sea surface, may not be able to detect birds flying at the top of the turbine RSZ \parencite{Schneider2024}. Telemetry studies can determine the location, flight height, and flight speed of tagged animals, offering the highest resolution information about an individual animal. However, tagged animals represent a small subset of the total population, and telemetry studies typically focus on a single species. Finally, radar systems can detect flying animals and determine their flight height and speed over ranges on the order of kilometers but typically can only classify animals into coarse size classes and currently cannot distinguish between birds and bats. 

Previous studies have highlighted that both wind speed and direction are important drivers of animal flight activity. \cite{Sjollema2014} noted a decrease in bat activity with increasing wind speeds. Both \cite{Ainley2015} and \cite{gibb2018} found that birds tended to change their flight styles (e.g., soaring versus flapping) at higher wind speeds, and \cite{kumagai2023} found Black-tailed Gulls (\textit{Larus crassirostris}) adjusted flight height and speed in response to changing wind conditions. In the North Sea, \cite{Bradaric2024} found that birds generally flew at altitudes with the most favorable wind conditions for their direction of travel (minimizing headwinds), and \cite{Skov2026} noted changes in micro-avoidance behavior around offshore wind turbines in response to wind conditions. \cite{Lagerveld2023} observed peaks in migratory movement of bats on nights with wind directions more aligned with travel direction (i.e., tailwinds). Similarly, \cite{Manola2020} found that during periods of migration fewer birds were present when there was a headwind in the direction of migration. However, several of these studies have largely relied on historical climate data with limited temporal and spatial resolution (e.g., the ERA5 reanalysis; \cite{Hersbach2020}) to analyze wind speed and direction, and none have analyzed animal abundance, flight height, and flight speed simultaneously. Further, most published studies investigating drivers of animal exposure to offshore wind turbines have taken place on the North Sea or other regions in Europe. In comparison, relatively few studies have been conducted along the east coast of the United States. Telemetry has been used to study the migratory patterns and habitat use of several key species in New England \parencite{Loring2014,Loring2020}, but, to the authors' knowledge, no studies to date have analyzed the influence of wind on animal behavior off the northeast coast of the United States. 

In this study, we present results from a co-spatial deployment of a $360^\circ$ radar system and two wind lidar systems on the Northeastern Shelf of the United States. These data were collected over a 5-week window in 2024 during autumn migration. The radar data contain information about the flight height and speed of travel of individual flying animals, and the lidar measurements provide high-resolution wind speed profiles. Because this work represents, to our knowledge, the first combined analysis of radar and lidar data to study the drivers of flying animal behavior, the methods presented in this paper provide a framework to integrate and analyze the two data types. The results offer insight into how wind influences animal abundance, flight direction, flight height, and flight speed during the autumn migration at this site. 

\section{Methods}\label{section:methods}

\begin{figure*}[b!]
    \centering
    \includegraphics[width=1\linewidth]{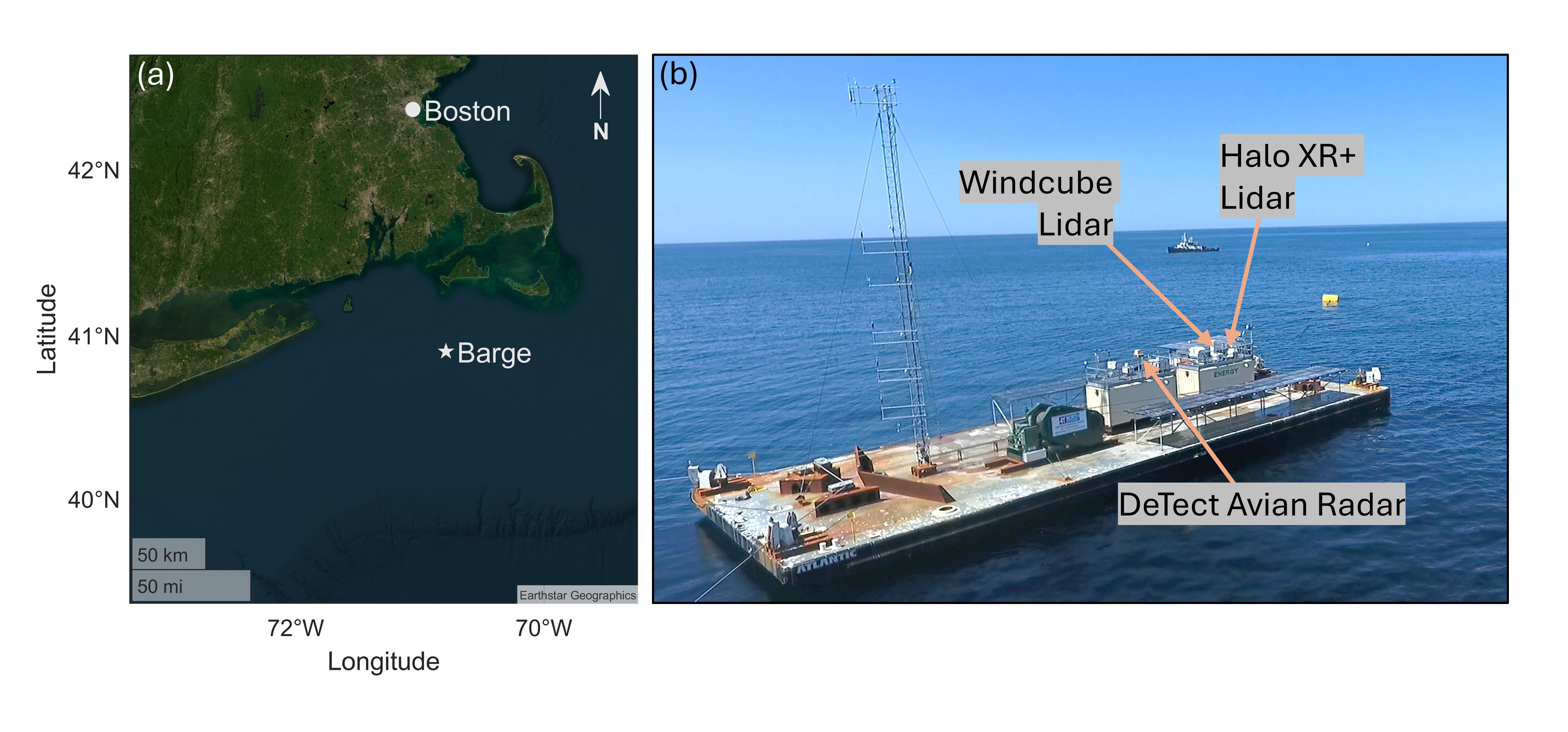}
    \caption{(a) Location of barge deployment south of Massachusetts, USA. Map was generated using the MATLAB satellite geobasemap, hosted by Esri. (b) Aerial photo of research barge, with locations of the DeTect avian radar and the two lidar systems used for wind measurements in this study. }
    \label{site}
\end{figure*}

In autumn 2024, a research barge (16.5 m x 61 m) was deployed as part of the Wind Forecast Improvement Project (\href{https://www2.whoi.edu/site/wfip3/}{WFIP-3}) off the northeast coast of the United States south of Massachusetts ($40.9^\circ$ N, $70.79^\circ$ W, Figure \ref{site}). This work considers bird and bat tracks from an avian radar (MERLIN True3D, DeTect, Panama City, Florida, USA) and concurrent wind measurements from two scanning lidars (WindCube v2.1, Vaisala, Vantaa, Finland, and Halo XR+, Halo Photonics, Lannion, France) installed on the barge. A meteorological mast (air pressure, temperature, rainfall) and a ceilometer (cloud ceiling altitude) were also installed on the barge, but data from these sensors were ultimately not used for this analysis. 

\begin{figure*}[t!]
    \centering
    \includegraphics[width=1\linewidth]{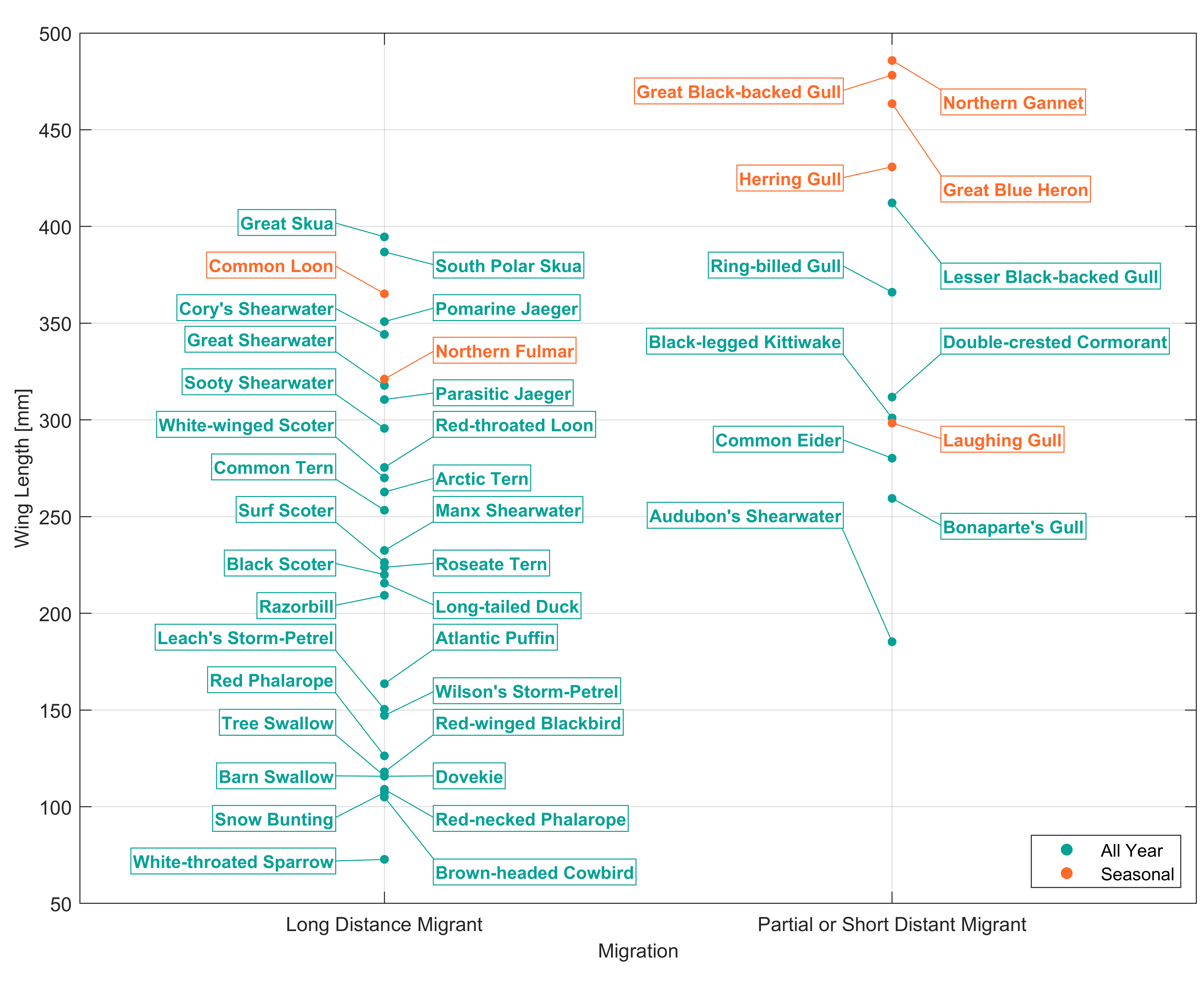}
    \caption{Species expected at the study location in autumn. Categorized by migration preference and seasonal versus year-round presence at the site. Migration preference and seasonal presence data are from \cite{Atlanticbirds} and expert opinion from the Biodiversity Research Institute. Wing length data are from \cite{avonet2022}.}
    \label{species}
\end{figure*}

Based on historical data, at least 43 bird species and three bat species can be expected to be present at the barge site during the data collection period \parencite{Atlanticbirds}. These species represent a range of morphologies and a mix of long-distance and short-distance migrants, with some species being partial migrants (Figure \ref{species}). Considering wing length (length from the carpal joint, bend of the wing, to the tip of the longest primary feather on the unflattened wing) as a coarse proxy for species size \parencite{avonet2022}, detected animals likely had typical wing lengths ranging between 72.8 mm (White-throated Sparrow) and 485 mm (Northern Gannet). 

\begin{figure*}[t!]
    \centering
    \includegraphics[width=1\linewidth]{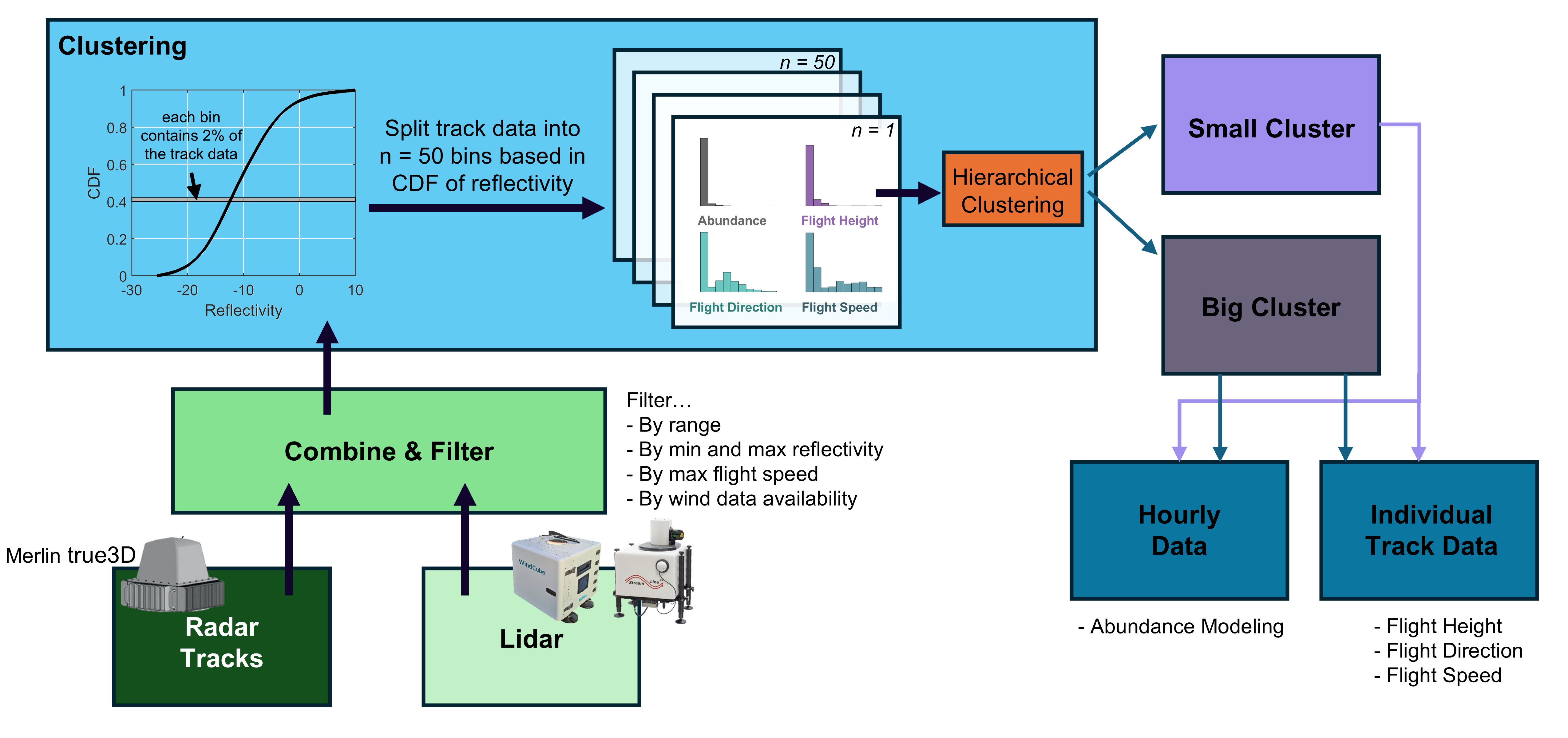}
    \caption{Data analysis flow chart}
    \label{clustering pipeline}
\end{figure*}

Radar and wind data were processed and aggregated before using a clustering algorithm to divide the track dataset into two groups based on approximate size (more detail on these groupings is presented in Section \ref{clustering}). A generalized additive model was developed to assess the drivers of hourly track abundance, and distributions of flight height, direction, and speed were investigated in relation to wind conditions. Figure \ref{clustering pipeline} provides an overview of the data processing and analysis, which are described in more detail in the subsequent sections.

\subsection{Radar Data}
The MERLIN True3D radar consists of four S-band pulsed Doppler panels with overlapping fields of view resulting in $360^\circ$ coverage. The radar was deployed from 16 June to 28 September 2024, but its internal inertial measurement unit did not function properly until 24 August, limiting motion compensation capabilities. The system was fully functional from 24 August to 28 September 2024. This time period forms the basis of our analysis. 

Bird and bat detections were automatically processed onboard the radar using proprietary tracking software developed by DeTect. The software output latitude, longitude, altitude above mean sea surface height, ground speed, and reflectivity for each detection in a track (a sequence of individual detections determined by the tracking software to represent a single animal). All post-processing of the track data took place in MATLAB and Python. 

For each track, speed, reflectivity, latitude, longitude, and altitude were calculated as the median of all detections comprising that track, and the flight direction was determined as the straight-line heading between the first and last detection in the track. To be consistent with the typical presentation of wind direction, flight direction is presented as the direction the track is flying from (e.g., if an animal was flying from 90$^\circ$ east to 270$^\circ$ west, its direction is 90$^\circ$). The body direction of the bird or bat may differ from the flight direction, but we do not attempt to determine body orientation in these data. 

Tracks were filtered by flight speed, reflectivity, range from the radar, and availability of wind data. First, tracks comprising fewer than five detections and/or median flight speeds greater than 35 m s\textsuperscript{-1} were removed from the dataset. Tracks that did not meet these criteria had limited information and/or had flight speeds above those for the fastest birds expected at the site like the Common Loon (\textit{Gavia immer
}, \cite{Alerstam2007}). Second, the dataset was limited to tracks with reflectivity between $-25.5$ and 10 dBsm. Tracks with median reflectivity values below $-25.5$ dBsm exhibited different behavior than tracks with higher reflectivity; their flight speeds were similar to the wind speed and their flight directions were dominantly to and from land. Based on these characteristics, we hypothesize that the majority of these small targets are insects and therefore are not representative of bird and bat behavior. Next, the dataset was filtered to only include tracks detected between 200 and 1100 m in range. This range was selected to capture the region of the sample volume where detection rates were relatively constant (a similar approach was employed in \cite{VanErp2021}). Analysis of the total number of detected tracks as a function of range revealed that due to the small sample volume at close ranges, the number of detections increased from the minimum detection range of 30 m to approximately 200 m, then remained relatively constant from 200 to 1100 m before dropping off as the detectability of smaller targets decreased. This range limit also served to filter out animals in close proximity to the barge that may have been influenced by the presence of the barge. Lastly, only tracks with co-temporal lidar measurements of wind speed and direction at flight height were analyzed (see Section \ref{wind methods}). 
 
While the radar system had a $360^\circ$ field of view, the number of detections varied with azimuth; the most detections were within the main lobes of the four antennas. More detail on spatial variability of detection rates is contained in Supplementary Info Figure 1. Because of the azimuthal variability in detection capabilities, this analysis only considers track distributions in altitude and does not consider distributions in latitude and longitude. Therefore, our analysis of trends in flight altitude requires the assumption that spatial variability in bird and bat behavior across the measurement area is small.

\subsection{Wind Data}
\label{wind methods}
Two lidars were installed on the barge to measure wind speed and direction. The first, a Windcube v2.1 profiling lidar, provided lower-altitude measurements from 40 m to 300 m in 20 m bins. The second, a Halo XR+ scanning lidar, provided higher-altitude measurements from 105 m to 2505 m in 30 m bins, but only data below 1500 m were considered since data quality significantly declined above this point. Both lidars performed velocity azimuth display scans to capture wind data. Filtered radial velocities \parencite{Krismanurthy2013} obtained from these scans were converted into wind speed and direction estimates using chi-square fitting techniques \parencite{Newsom2022}. The wind profiles were adjusted to Earth-fixed coordinates using the inertial measurement unit on the barge \parencite{Krismanurthy2023}. Data from both lidars were scalar-averaged into 10-minute intervals and linearly interpolated to a single grid with 30 m altitude bins. This combined dataset contains mean wind speed, wind direction, and turbulence intensity (TI) in 10-minute time bins and 30 m altitude bins. To facilitate alignment with track flight heights, these data were linearly interpolated to 10 m altitude bins in MATLAB (Supplementary Info Figure 2).  

Wind data were used to determine the wind speed and direction at flight height and to calculate wind assistance, air speed, and altitude of maximum wind assistance for each track. Wind speed and direction for each track were determined using the altitude bin closest to flight height. If wind data were not available within 30 m of altitude or 10 minutes of the track start time (the spatial --before interpolation-- and temporal resolution of the lidar dataset), the track was removed from the dataset. Each track was classified as traveling with a tailwind, crosswind, or headwind based on the difference between flight and wind direction at flight height (tailwinds: $-60^\circ$ to $60^\circ$, crosswinds: $-120^\circ$ to $-60^\circ$ or $60^\circ$ to $120^\circ$, headwinds: $<-120^\circ$ or $>120^\circ$).   

Wind assistance is the component of the wind in the travel direction of a bird or bat and can be thought of as the portion of the wind that aids or impedes flight \parencite{kumagai2023}. Wind assistance as a function of altitude, $z$, for each track was computed as
\begin{equation}
WA(z) = v(z)cos(\theta(z)),
\end{equation}
where $v$ is the velocity magnitude of the wind at the closest 10-minute average to the track start time stamp, and $\theta$ is the difference between the track and wind directions. Air speed for each track (flight speed relative to the wind) was calculated by subtracting wind assistance at flight height from the measured ground speed. The altitude of maximum wind assistance (i.e., the altitude where the animal would experience the most wind assistance given the time when it was detected and its measured flight direction) was determined by calculating theoretical wind assistance at every altitude at the time of detection. Tracks were removed from the dataset if the altitude of maximum wind assistance corresponded to the lowest bin of available wind data and wind data were missing at lower altitudes or, conversely, if the altitude of maximum wind assistance corresponded to the highest bin of available wind data. Wind availability at flight height and the altitude of maximum wind assistance filters removed 21.19\% of tracks from the dataset (Supplementary Info Figure 3). A large portion of the removed tracks (36.1\%) occurred between 11 September and 16 September when the wind data were most intermittent. 

Cumulative distribution plots of flight direction and altitude are presented as being conditional on either mean wind speed or hour of the day. When conditional on mean wind speed, the track dataset was partitioned into 10\% bins based on the profile mean of the wind speed (mean across altitude) at the time stamp closest to the track start time. When conditional on hour of the day, the tracks were partitioned by the start time of the track, with hour $0$ corresponding to 00:00--00:59 local time (Eastern Daylight Time). After partitioning, the cumulative distribution function was plotted for each bin using the \emph{cdfplot} function in MATLAB.

\subsection{Sun Azimuth and Elevation}
Sun azimuth and elevation angles were computed at the start of each hour for the barge deployment location (with \cite{suncode}). Sun azimuth is referenced to 0$^\circ$ = north, and negative sun elevation angles denote positions below the horizon. Sun azimuth and elevation were used as predictors for abundance modeling (Section \ref{sec:abundancemodelmethods}) and to consider track abundance during four distinct times of day: dawn, day, dusk, and night. Dawn and dusk are defined as the hours encompassing astronomical, nautical, and civil twilight (sun elevation between $-18^\circ$ and $0^\circ$); dawn corresponds to the period before sunrise and dusk corresponds to the period after sunset. Day is defined as the hours between sunrise and sunset, and night is defined as all hours between dusk and dawn. The exact hour of the day demarcating the different day periods changed throughout the study window. 

\subsection{Clustering}
\label{clustering}
A key limitation of avian radar data is the inability to determine species or animal size. Ideally, the data could be grouped in some way to study more granular trends that are otherwise difficult to parse because of large variability in the full dataset. Here, we partition the data based on approximate size groups using reflectivity as a proxy for animal size. Reflectivity of an individual target (i.e., the strength of the returned radar signal) generally increases with animal size, but also depends on the liquid content and orientation of the animal. For example, the same animal flying head-on would likely have lower reflectivity than if it were broadside to the radar beam. Therefore, it is difficult to determine radar reflectivity thresholds that separate the data into meaningful groupings for analysis. To address this challenge, we used a hierarchical clustering approach to split the data into two approximate size groups that exhibited similar behaviors (Figure \ref{clustering pipeline}). 

To perform clustering, track data were first subdivided into 2\% bins based on the CDF of track reflectivity (i.e. $n = 50$ total bins that each contain 2\% of the track dataset). Then, the \emph{clusterdata} function in MATLAB was used to sort these reflectivity bins into two clusters. Two clusters were chosen because the resulting clusters delineated the data well (see more details in Section \ref{Results}) and enabled tractable analysis. The clustering inputs for each reflectivity bin were all possible combinations of four probability histograms for the tracks in the bin: hourly abundance (number of tracks per hour of the year), flight height, flight direction, and flight speed. The clustering divided the data into one cluster with higher reflectivity values, and one cluster with lower reflectivity values. For each combination of inputs, the euclidean distance between the two clusters in the four-dimensional space spanned by the input variables as well as the reflectivity value that divides the two clusters are given in Table \ref{clustering table}. 

The best performing clustering implementations were determined to be the combinations of inputs that maximized the euclidean distance between clusters (i.e., the two clusters were the most distinct from eachother). Three implementations tied (cluster distance equal to 0.3704) and resulted in the same reflectivity divide ($-8.54$ dBsm). All three of these included flight height as an input. Of these, the implementation using flight height, flight direction, and flight speed is carried forward for the rest of the analysis. While clustering was performed on reflectivity bins, all subsequent analysis is based on the individual tracks. Each track is assigned to the cluster that its reflectivity bin was placed in: ``small'' for median track reflectivity values less than or equal to the cluster divide value of $-8.54$ dBsm, and ``big'' for values greater than $-8.54$ dBsm. 

\begin{table}[t]
\centering
\setlength{\tabcolsep}{4pt}
\caption{Clustering input combinations and resulting euclidean distance and reflectivity divide between clusters. The best performing implementations (based on euclidean distance) are in bold.}
    \begin{tabular}{c c c c c c}
    \toprule
        Hourly Abund. & Flight Height  & Flight Dir. & Flight Speed & Cluster Dist. & Reflectivity Divide [dBsm] \\
    \midrule
    X    & X & X & X & 0.3551 & $-10.21$  \\
    X & X & X &  & 0.3551 & $-10.21$  \\
    X    & X &   & X & 0.3551 & $-10.21$  \\
    X & X &  &  & 0.3659 & $-9.23$  \\
    X &  & X & X & 0.2401 & $-20.16$  \\
    X &  & X &  & 0.3551 & $-10.21$  \\
    X &  &  & X & 0.2401 & $-20.16$  \\
    X &  &  &  & 0.3471 & $-10.84$  \\
    & \textbf{X} & \textbf{X} & \textbf{X} & \textbf{0.3704} & $\mathbf{-8.54}$  \\
    & \textbf{X} & \textbf{X} &  & \textbf{0.3704} & $\mathbf{-8.54}$  \\
    & \textbf{X} &  & \textbf{X} & \textbf{0.3704} & $\mathbf{-8.54}$  \\
    & X &  &  & 0.341 & $-11.45$  \\
    &  &  X & X & 0.1333 & $-16.605$  \\
    &  & X &  & 0.3551 & $-10.21$  \\
    &  &  & X & 0.1787 & $-20.16$  \\

    \bottomrule
    \end{tabular}
    \label{clustering table}
\end{table}

\subsection{Abundance Modeling}\label{sec:abundancemodelmethods}
Generalized additive models (GAMs) of hourly abundance (number of tracks per hour of the year) were used to investigate the meteorological and astronomical drivers of bird and bat presence. Two GAMs were fit, one for each cluster. The developed models are not meant to be predictive, and no effort is made to investigate predictive accuracy. Instead, the goal is to use the abundance models to estimate the individual effects of predictor variables on abundance in this 5-week study window. Modeling was implemented with the \emph{pyGAM} package in Python with a Poisson distribution and a log link function. Model predictors were down-selected from a larger list based on data availability and to ensure all included predictors had variance inflation factors less than 4 (a check for multicollinearity). Removed predictors include air pressure and temperature, cloud ceiling altitude, and rainfall. Lunar illumination was also considered but ultimately not included in the model because the 5-week dataset did not include multiple lunar cycles. Wind speed and direction were used as inputs instead of wind assistance (a function of both terms) to consider the effects of each parameter separately. 

The final predictor set included hourly averages of wind speed, direction, and turbulence intensity (averaged across all 10-minute averages in the hour and across all altitude bins), as well as sun elevation and azimuth. Additionally, abundance from the previous hour, ``animals prior,'' was included as a predictor to account for autocorrelation of the model residuals \parencite{VanErp2021}. The pyGAM \emph{gridsearch} method was used to choose the best smoothing hyperparameter, $\lambda$, by minimizing the generalized cross-validation score. The final models used $\lambda\:=\:1000$ and splines with 20 basis functions.

The importance of each predictor to hourly abundance is quantified by the percent deviance explained in the model by the predictor. Following the methods used in \cite{VanErp2021}, this was approximated by subtracting the deviance explained by a sub-model (model without the predictor) from the deviance explained by the full model, then dividing by the deviance explained by the full model. This is a representation of the predictor's individual effect, but does not account for shared deviance between predictors \parencite{Lai2024}.

\subsection{Relating Results to Wind Turbine Operation}
\label{turbrelate}
While offshore wind turbines vary in size and rated power, it is instructive to consider a representative wind turbine to place results in context. Here, we consider the International Energy Agency (IEA) 15 MW reference offshore wind turbine \parencite{RefTurb}. This wind turbine has a 240 m rotor diameter and a hub height of 150 m, meaning that the RSZ spans 30 to 270 m in altitude. 

Wind turbines are typically controlled relative to wind direction and speed. Optimal power generation occurs if the turbine rotor is perpendicular to the prevailing wind direction, so wind turbines actively yaw (rotate) to align the RSZ with the incoming wind as conditions change. This means that animals flying with headwinds or tailwinds would be traveling oblique to the RSZ, and animals flying with crosswinds would fly parallel to the RSZ. Additionally, wind turbines typically operate in four distinct control regions based on the prevailing wind speed. In Region I, wind speeds are slow enough that it is not economical to operate wind turbines, so they are stalled. Region II represents wind speeds where turbines are operating and the rotor rotation rate increases with wind speed to maximize efficiency. Beyond a rated wind speed, wind turbines enter Region III, in which the control objective is to keep power constant at the rated power using a combination of acceleration, deceleration, and/or pitch control under increasing wind speeds. In Region IV, operation is curtailed when wind speeds exceed a defined cutout speed to avoid damage to the turbine. Region II and Region III are most relevant to collision risk with moving structures because the turbine is rotating in these operational states. The region boundaries for the IEA 15 MW reference turbine are as follows: Region 1, 0--3 m s\textsuperscript{-1}; Region II, 3--11 m s\textsuperscript{-1}; Region III, 11--25 m s\textsuperscript{-1}; Region IV, $>$25 m s\textsuperscript{-1}.

\section{Results}
\label{Results}

\begin{figure*}[t!]
    \centering
    \includegraphics[width=1\linewidth]{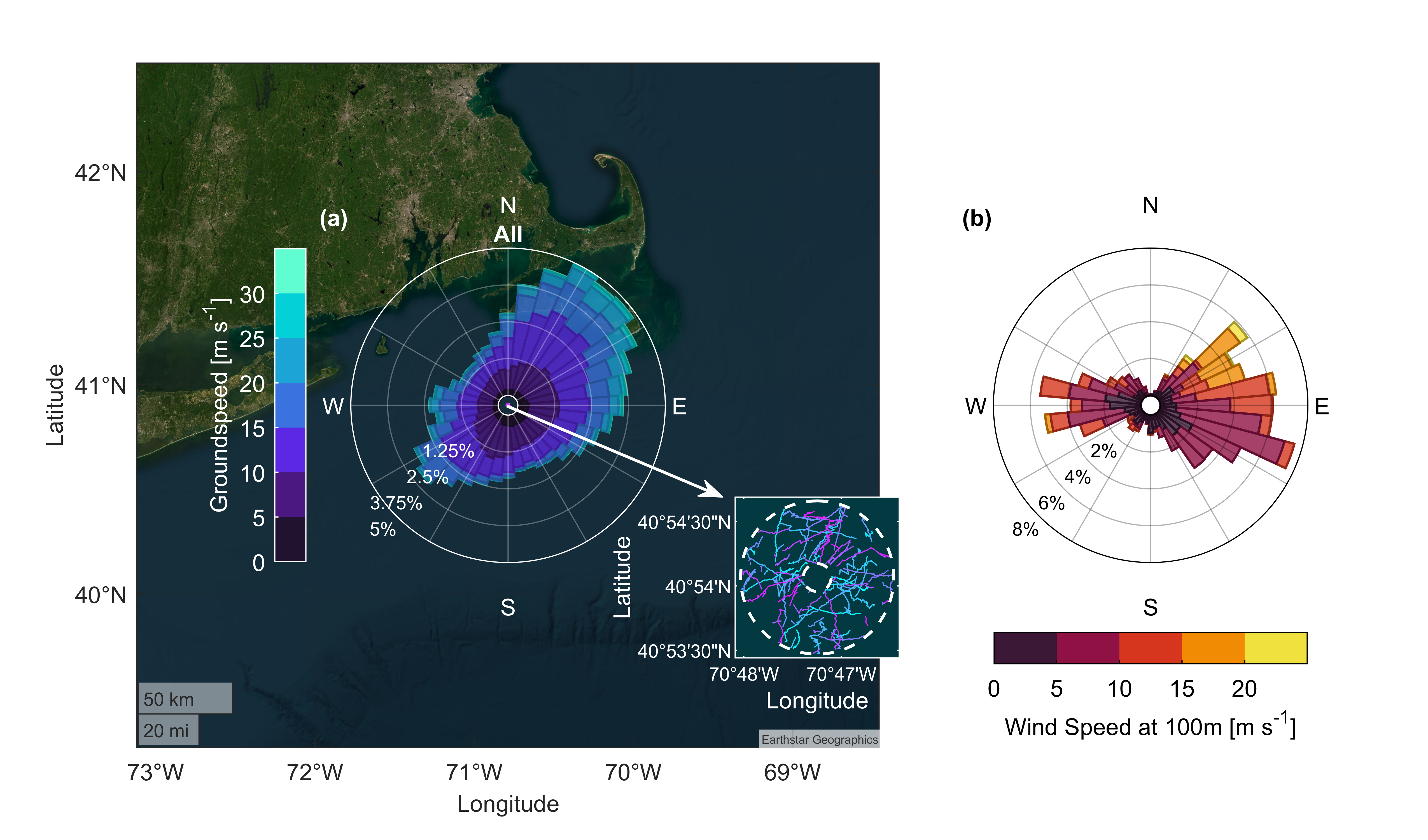}
    \caption{(a) Flight speed and direction rose for all tracks. Note: flight direction is the direction the tracks are originating from. The inset highlights 100 randomly selected tracks from the radar dataset; the white dashed lines represent the range limits applied in post-processing. (b) Wind speed and direction rose at 100 m altitude for the 5-week study window.}
    \label{site overview}
\end{figure*}

\begin{figure*}[t!]
    \centering
    \includegraphics[width=1\linewidth]{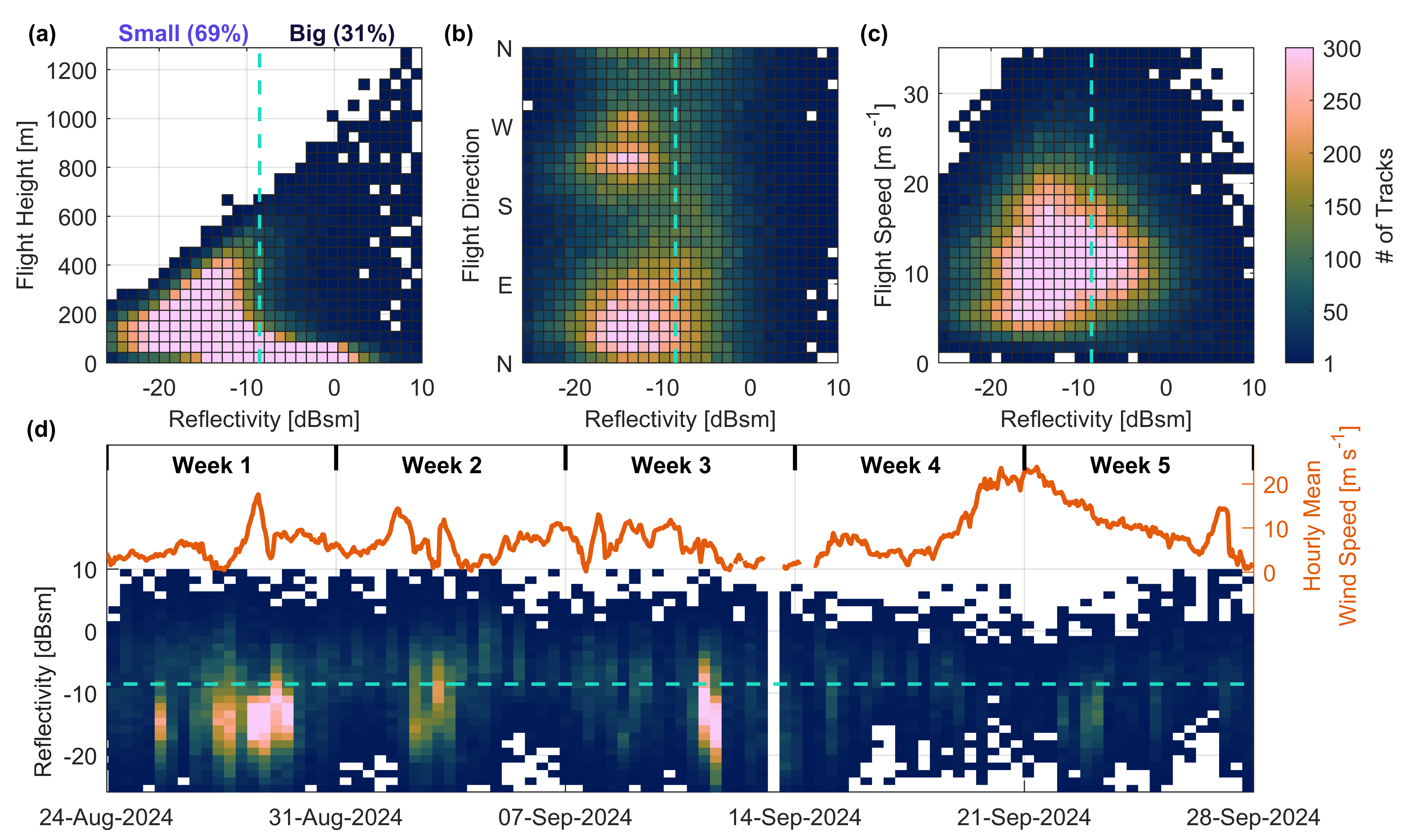}
    \caption{Bivariate histograms of reflectivity with (a) flight height, (b) flight direction, (c) flight speed, and (d) date. The cyan dashed line represents the reflectivity delineation between the big and small size clusters. Note: flight direction is the
    direction the tracks are originating from  }
    \label{heatmaps}
\end{figure*}

After all filtering steps, 67,410 tracks remained out of 301,618. Figure \ref{site overview}a shows a flight speed and direction rose for the remaining tracks with the site geography for context. Most tracks (88.9\%) have speeds between 5 and 25 m s\textsuperscript{-1}. The dominant flight direction follows expectations for a southern autumn migration with the majority of tracks coming from the northeast. Based on these bearings, birds and bats may be originating from southeastern Massachusetts. A smaller but still substantial number of animals traveled from the southwest. Figure \ref{site overview}b shows a wind rose of wind conditions at 100 m altitude at every time step over the entire study window. Winds were most common out of the east, were less common from the west, and rarely blew from due north or due south over the study window. The strongest winds occurred out of the northeast. 

Figure \ref{heatmaps} shows bivariate histograms of track reflectivity as a function of (a) flight height, (b) flight direction, (c) flight speed and (d) date. The divide between the two clusters ($-8.54$ dBsm) is indicated in each heatmap, and the time series of mean wind speed is shown for reference. 69\% of tracks were assigned to the small cluster and 31\% to the big cluster. The heatmaps reveal that the reflectivity threshold identified through the clustering analysis separates distinct trends in flight behavior. Notably, animals in the small cluster tended to fly at higher altitudes and over a broader range of altitudes, whereas animals in the big cluster were more confined below 100 m (Figure \ref{heatmaps}a). The majority of animals in the small cluster flew from the northeast or southwest; there is not a clear trend in flight direction for big targets (Figure \ref{heatmaps}b). Conversely, a broader spread of flight speeds is observed for small targets than for big targets (Figure \ref{heatmaps}c), with a hotspot in flight speed for big targets between 8 and 14 m s\textsuperscript{-1}. Lastly, Figure \ref{heatmaps}d reveals temporal bursts in the presence of smaller animals while the larger animals had a more consistent baseline of activity. The hour with the highest number of detected tracks occurred on 11 September 2024 with 1007 tracks from the small cluster and 68 tracks from the big cluster. Across the 5-week study window, hourly mean wind speeds rarely exceeded 15 m s\textsuperscript{-1}, with the exception of a strong wind event bridging weeks 4 and 5 when mean wind speeds exceeded 15 m s\textsuperscript{-1} for a 36-hour period (Figure \ref{heatmaps}d). There were fewer detections during this period. Hourly mean wind speeds did not exceed 25 m s\textsuperscript{-1}, the Region IV cutout speed, at any point during the study period. While these heatmaps mask some interesting trends in the data due to the constraints of a color map, we have included them to provide a high-level overview of the dataset. The remainder of this section provides more in-depth analysis of abundance, flight direction, flight height, and flight speed.

\begin{figure*}[t!]
    \centering
    \includegraphics[width=1\linewidth]{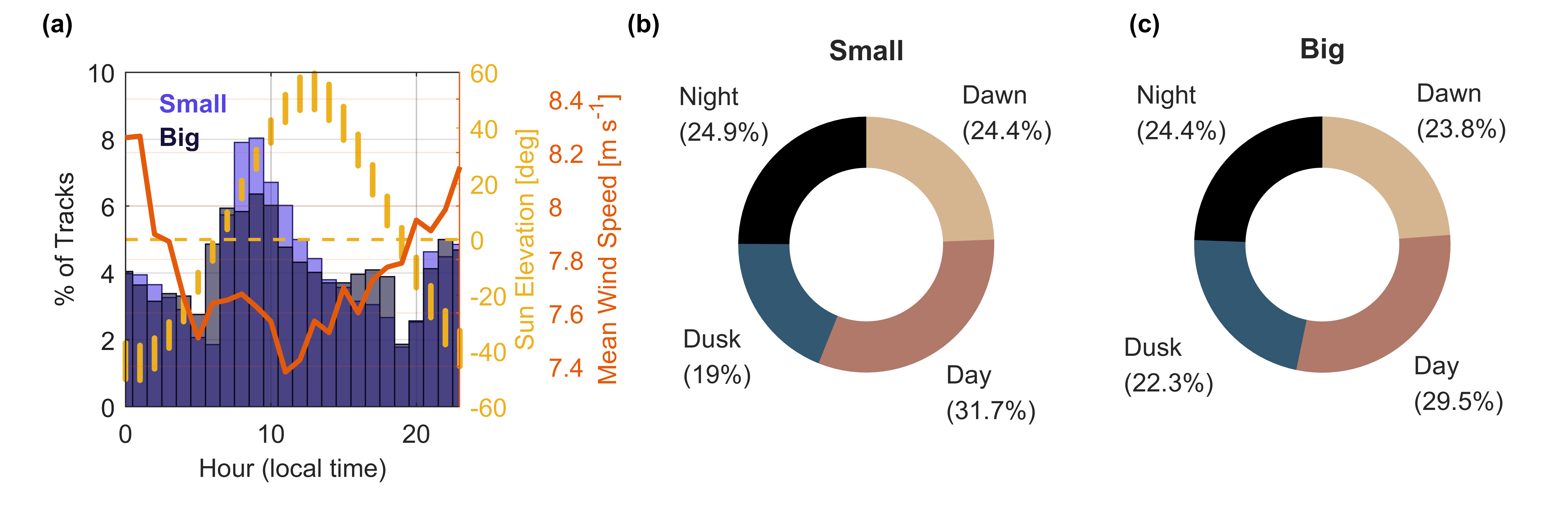}
    \caption{(a) Histogram of small and big cluster tracks with respect to hour of the day. The range of sun elevations observed during each hour of the day and the mean wind speed over the study window are shown for reference. Abundance proportion for each day period is standardized by the number of hours in a given period of the day for the (b) small and (c) big clusters.}
    \label{abundance overview}
\end{figure*}

\subsection{Abundance}
Figure \ref{abundance overview} provides an overview of trends in the number of track detections with hour of the day. For both size clusters, there is a peak in detections in the hours immediately following sunrise. After sunrise, the number of detections generally decreased throughout the day before reaching a secondary peak at night (Figure \ref{abundance overview}a). While this trend is consistent between the two size clusters, the morning peak in detections is more pronounced for the small cluster. This peak in abundance corresponds to the hours of the day where wind speeds are typically lowest. Figure \ref{abundance overview}b and c show that detections are relatively evenly split between periods of the day (dawn, day, dusk, night) when standardized by the number of hours in a given period of the day. For both size clusters, a slightly higher proportion of detections occurred during the day (31.7\% of small cluster tracks; 29.5\% of big cluster tracks) and a slightly lower proportion occurred at dusk (19\% of small cluster tracks; 22.3\% of big cluster tracks).

\begin{figure*}[t!]
    \centering
    \includegraphics[width=1\linewidth]{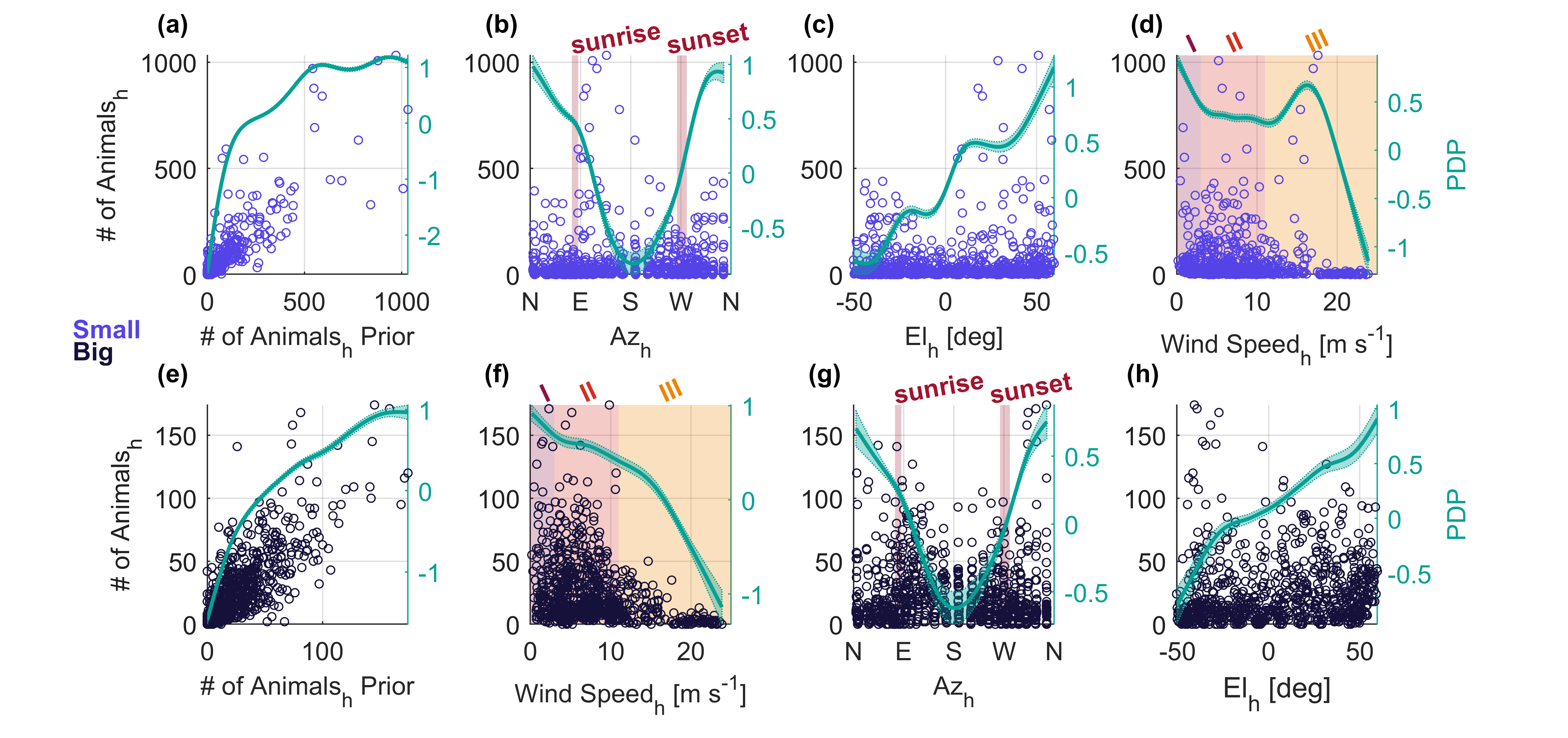}
    \caption{Partial dependence plots (PDP) for the top four predictors in the GAM abundance models for the (a--d) small cluster and the (e--h) big cluster. Shaded cyan region represents the 95\% confidence interval. Open circles represent measured hourly abundance. Wind turbine control regions I - III (Section \ref{turbrelate}) are denoted in (d) and (f).}
    \label{PDP}
\end{figure*}

The GAM modeling results are presented in Table \ref{modeling results} and Figure \ref{PDP}. The model fits explain the majority of the deviance in each cluster dataset. The fit is better for the small cluster (85.5\% deviance explained), than for the big cluster (67.8\% deviance explained). For both models, animals prior (abundance during the previous hour) accounted for the highest proportion of the deviance explained, suggesting that it is the strongest predictor of hourly abundance. Of the environmental predictors included in the final model, wind speed and sun azimuth were the most impactful in the big cluster model, while sun azimuth, sun elevation, and wind speed were the most impactful in the small cluster model. Specifically, wind speed and sun azimuth respectively account for 2.9\% and 2.1\% of the deviance explained by big cluster model. For the small cluster model, sun azimuth accounts for 1.5\% of the deviance explained, and sun elevation and wind speed each account for 1.2\% of the deviance explained. Wind direction and turbulence intensity were the lowest ranked predictors for both clusters. 

\begin{table}[t!]
\centering
\setlength{\tabcolsep}{4pt}
\caption{Deviance explained by the GAM for each cluster (``Total'') and the estimated percent deviance explained by each predictor.}
    \begin{tabular}{c c c c c c c c}
    \toprule
       & Total & Animals Prior& Wind Speed  & Wind Direction & TI & Sun Azimuth & Sun Elevation \\
    \midrule
    Small    & 85.5\% & 48.5\% & 1.2\% & 0.26\% & 0.4\% & 1.5\% & 1.2\%  \\
    Big    & 67.8\% & 36.5\% & 2.9\% & 0.77\% & 0.25\% & 2\% & 1.4\%   \\
    
    \bottomrule
    \end{tabular}
    \label{modeling results}
\end{table}

The partial dependence plots in Figure \ref{PDP} highlight the average individual effect of the top four predictors of hourly abundance for each model. Positive values on the right-hand $y$-axis mean the model fit indicates that the condition on the $x$-axis positively influences abundance, and a negative value indicates that the condition suppresses abundance. The models for both clusters show an increase in abundance as animals prior increases (Figure \ref{PDP}a,e). For the small cluster, Region I and Region III wind speeds (up to 20 m s\textsuperscript{-1}) most positively impacted abundance, whereas wind speeds in Region II had a more marginal effect, and wind speeds above 20 m s\textsuperscript{-1} negatively impacted abundance (Figure \ref{PDP}c). The most positive association between wind speed and abundance for the big cluster occurs for Region I wind speeds before dropping off continuously. Wind speeds above 15 m s\textsuperscript{-1} have a negative effect on abundance for the big cluster. Wind speeds above 20 m s\textsuperscript{-1} negatively affected abundance regardless of cluster, and extrapolation would suggest that this trend persists in Region IV wind speeds that were not observed during the study period. The sun azimuth and elevation partial dependence plots are similar between both size clusters. Specifically, sun azimuths corresponding to the hours before sunrise and after sunset are associated with increased abundance while lower abundance is associated with midday (Figure \ref{PDP}c,g). Further, sun elevations above $0^\circ$ (daytime) positively influence abundance while elevations below $0^\circ$ negatively influence abundance (Figure \ref{PDP}d,h). 

\subsection{Flight Direction}

\begin{figure*}[t!]
    \centering
    \includegraphics[width=1\linewidth]{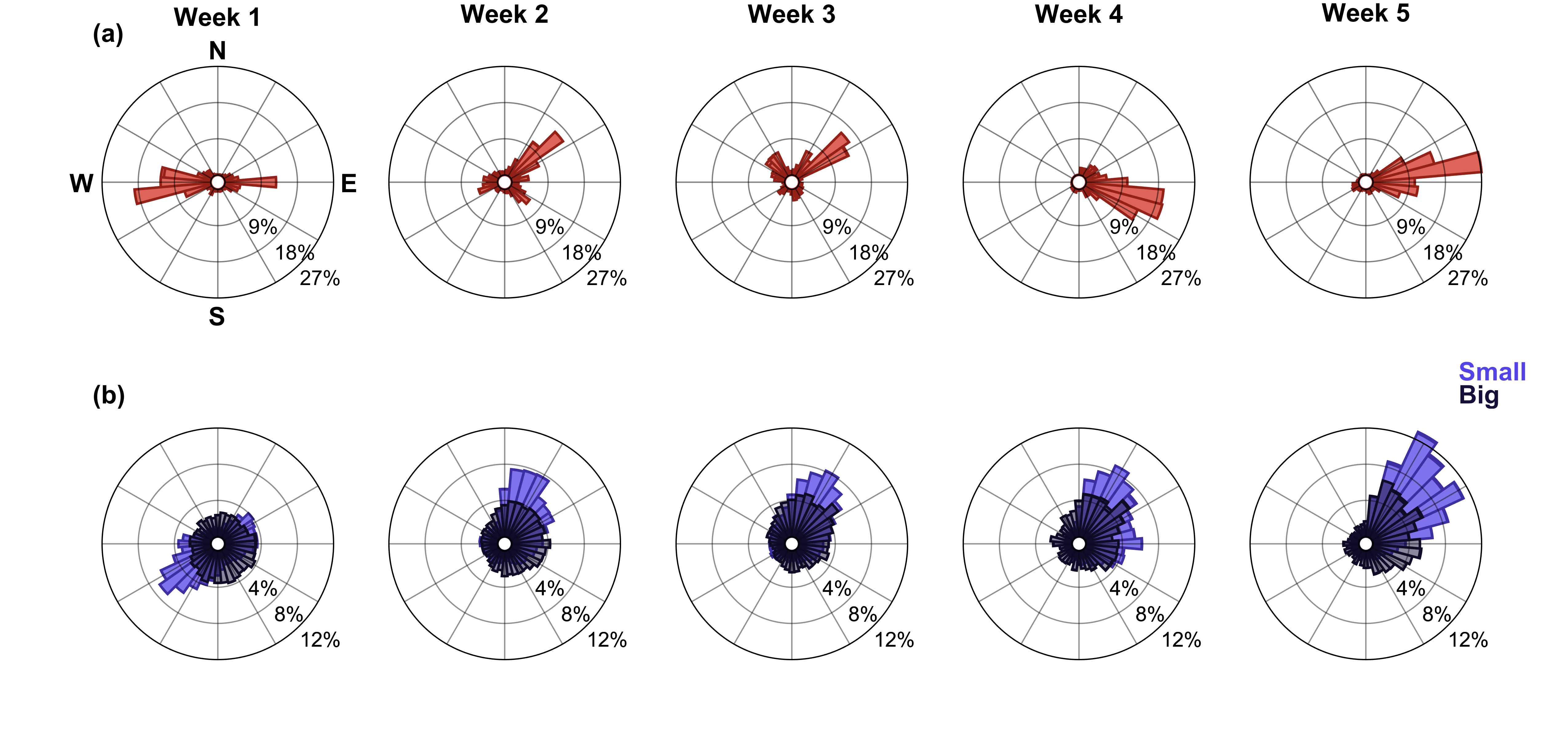}
    \caption{(a) Weekly wind speed and direction polars for wind conditions at flight height. (b) Weekly flight speed and direction polars for the two size clusters. Note: flight direction is the direction the tracks are originating from}
    \label{weekly wind and flight polars}
\end{figure*}

\begin{figure*}[t!]
    \centering
    \includegraphics[width=1\linewidth]{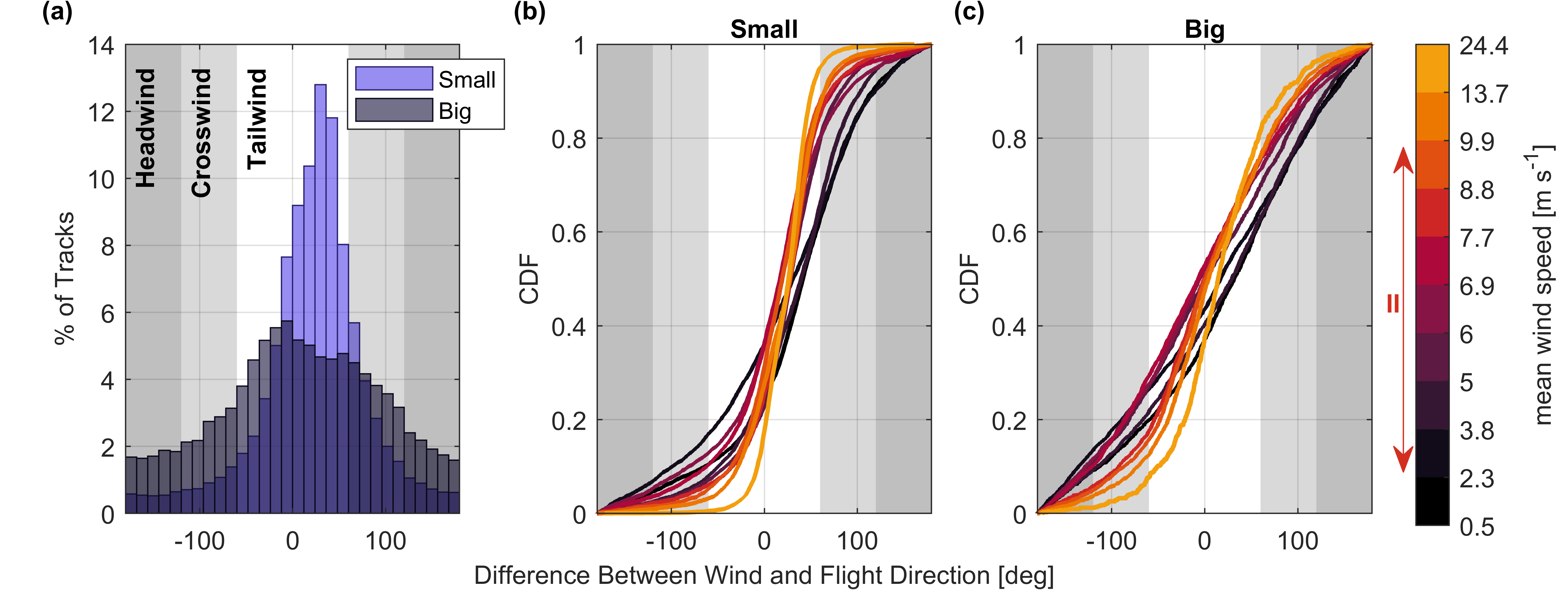}
    \caption{(a) Histograms of difference between wind and flight direction for the two size clusters. Cumulative distribution function (CDF) of difference between wind and flight direction conditional on mean wind speed for (b) the small cluster and (c) the big cluster. Region II wind speeds are denoted on the colorbar.}
    \label{conditional flight direction}
\end{figure*}

\begin{figure*}[t!]
    \centering
    \includegraphics[width=1\linewidth]{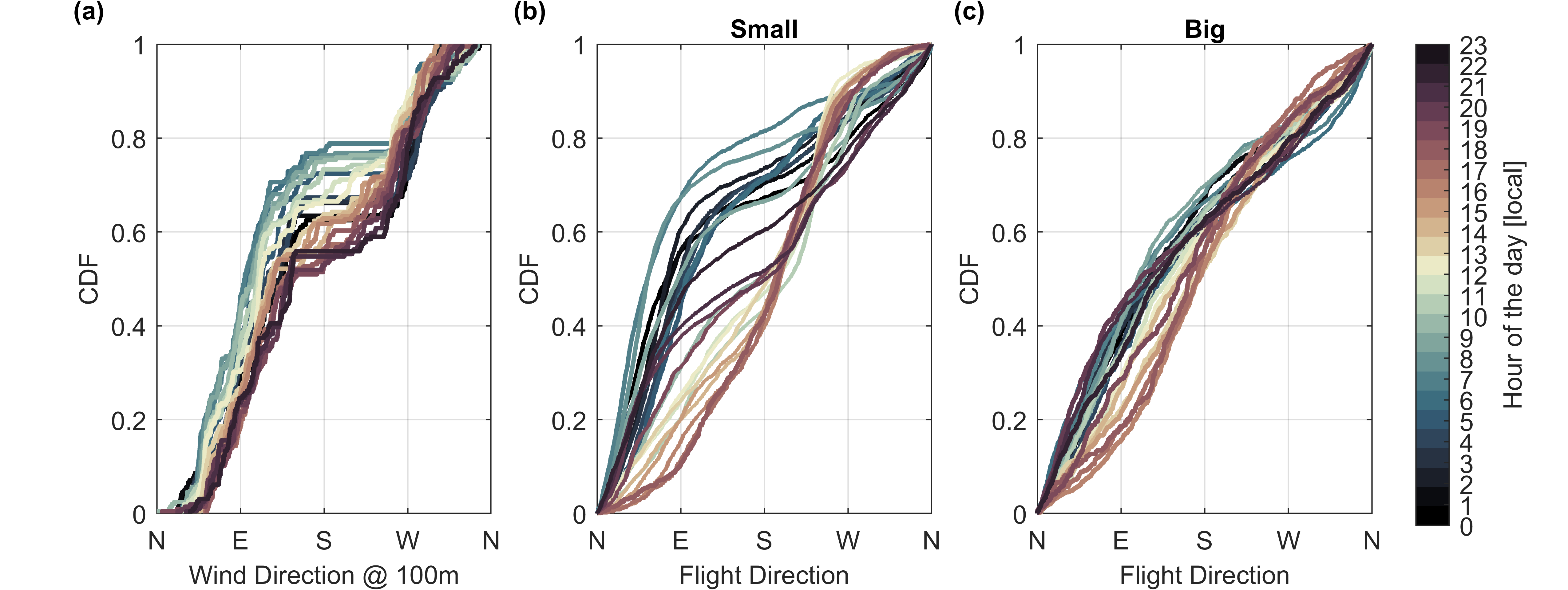}
    \caption{Cumulative distributions conditional on hour of the day for (a) wind direction at 100 m, (b) small cluster flight direction, and (c) big cluster flight direction. Note: flight direction is the direction the tracks are originating from}
    \label{conditional flight direction hour of day}
\end{figure*}


Wind and flight direction distributions shifted throughout the study window. Figure \ref{weekly wind and flight polars} shows polar plots of the flight direction for each track and corresponding wind speed and direction at flight height by week of the study window. Similar trends are apparent between the wind direction at flight height and flight direction. In Week 1, the wind direction at flight height is dominantly from the west, but by Week 2, and through the rest of the study window, it is dominantly out of the east. As the wind direction shifts, so do the dominant flight directions. This trend is present for both size clusters but is most apparent for tracks in the small cluster. For instance, tracks in the small cluster were predominantly out of the southwest in Week 1 and predominantly out of the northeast in Week 5. Further, tracks in the small cluster had more concentrated flight directions than the big cluster, but the flight direction distribution for tracks in the big cluster narrowed in the later weeks of the study window.

To further analyze the relationship between wind and flight direction, Figure \ref{conditional flight direction} highlights trends in the difference between flight direction and the wind direction for each track. The histograms for both clusters in Figure \ref{conditional flight direction}a indicate the majority of animals flew with tailwinds (i.e., difference between flight and wind direction near zero). However, this trend is less pronounced for tracks in the big cluster, indicating that larger animals were more likely than the smaller animals to fly with unfavorable winds. Specifically, the majority of animals in the small cluster (72.4\%) flew with a tailwind while 49.1\% of those in the big cluster did. Of the small cluster, 20.9\% of animals flew in crosswinds and only 6.8\% in headwinds. Of the big cluster, 32.3\% of animals flew in crosswinds and 18.6\% in headwinds. Cumulative distributions of the difference between wind and flight direction conditional on mean wind speed are shown in Figure \ref{conditional flight direction}b,c. At higher wind speeds, the distributions for both size clusters narrow, with a higher percentage of animals flying with tailwinds. For example 42.6\% of big cluster animals and 52.5\% of small cluster tracks flew with tailwinds in mean wind speeds between 0.5 and 2.3 m s\textsuperscript{-1} compared to 73.8\% of the big cluster and 94.5\% of the small cluster in wind speeds between 13.7 and 24.4 m s\textsuperscript{-1}. 

Flight directions for tracks in the small cluster exhibit diurnal variation similar to the diurnal variation in the wind direction. Figure \ref{conditional flight direction hour of day} shows the cumulative distribution of wind and flight direction conditional on hour of the day. For hours in the first half of the day (cool tones), easterly winds are most common, while in the later half of the day (warm tones), the distribution shifts towards more westerly winds. For example, 77.3\% of wind measurements at 100 m were from the east ($0^\circ$--$180^\circ$) and 22.7\% from the west ($180^\circ$--$360^\circ$) at 07:00, versus 61.7\% and 38.3\%, respectively, at 17:00. A similar, and more exaggerated, pattern is evident in the distributions of tracks in the small cluster, but this trend is less evident in flight directions of tracks in the big cluster. Specifically, 81.5\% of small cluster tracks at 07:00 were from the east and 18.5\% from the west. At 17:00 these proportions shift to 39.9\% and 60.1\%, respectively. In comparison, 66.5\% of tracks in the big cluster were from the east at 07:00 and 52.7\% at 17:00. 


\begin{figure*}[t!]
    \centering
    \includegraphics[width=1\linewidth]{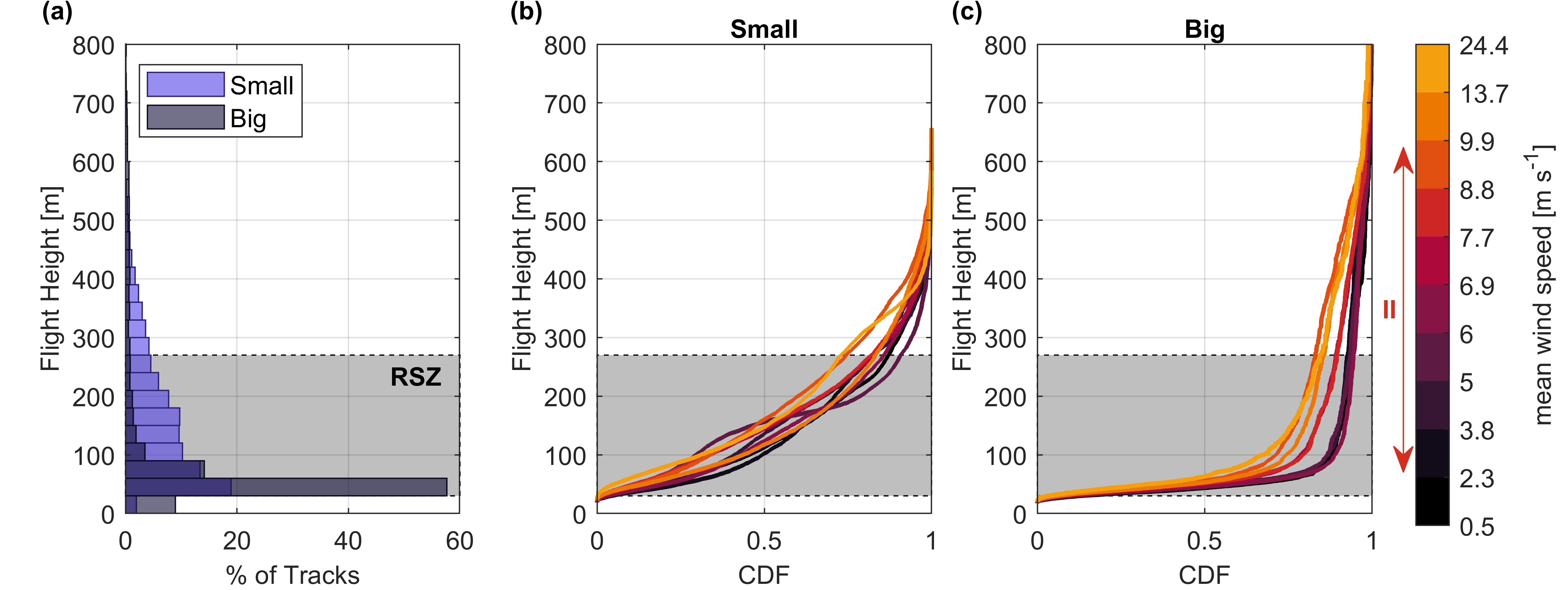}
    \caption{(a) Flight height distribution histograms for each size cluster. (b,c) Cumulative distribution function for flight height conditional on mean wind speed for the small cluster and the big cluster. Region II wind speeds are denoted on the colorbar. Note: CDF is on the x-axis in (b) and (c).}
    \label{flight height stationarity}
\end{figure*}

\begin{figure*}[t!]
    \centering
    \includegraphics[width=1\linewidth]{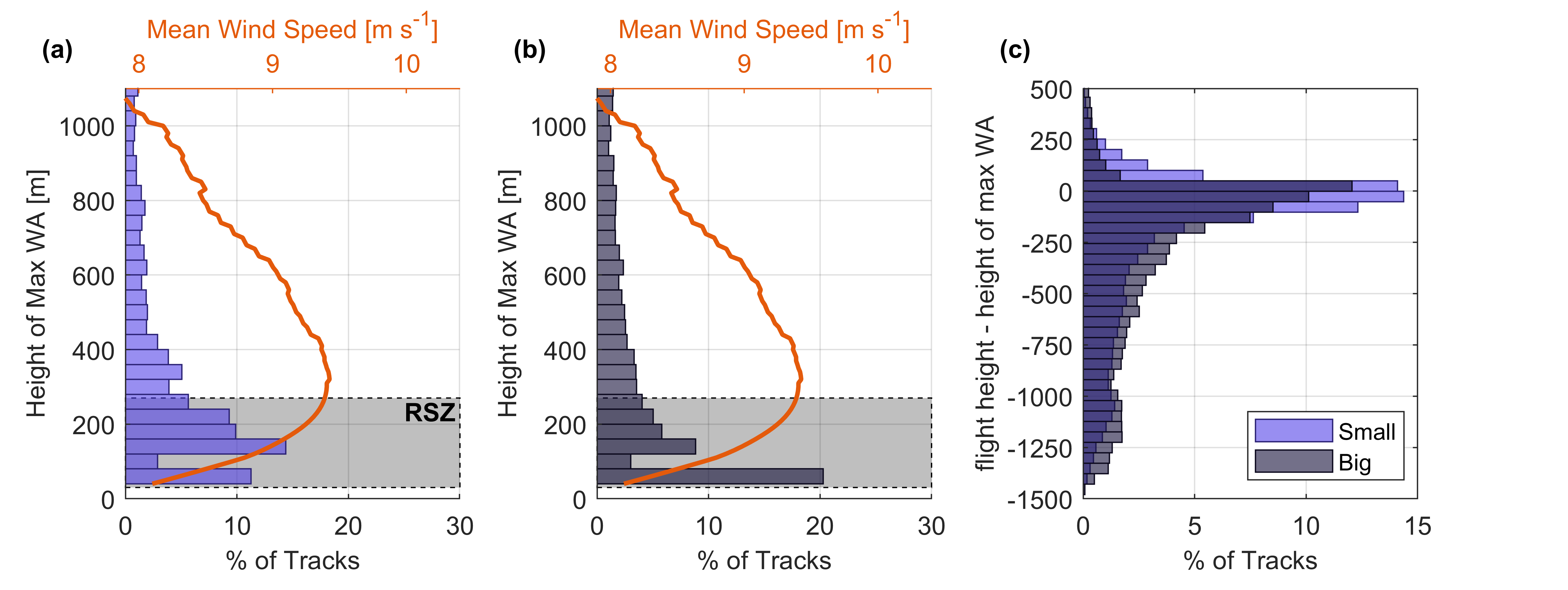}
    \caption{(a,b) Height of maximum wind assistance (WA) histogram for the small cluster and the big cluster. The orange line indicates the mean wind speed profile across the study window. (c) Histogram of the difference between track flight height and the height of maximum wind assistance by size cluster.}
    \label{Wind assist}
\end{figure*}

\subsection{Flight Height}

Figure \ref{flight height stationarity} shows the distributions of flight height for all tracks in each cluster. Most tracks (80.6\% of all tracks) were within the reference turbine RSZ. However, the minimal number of detections below the RSZ could be a consequence of limited radar detection capabilities near the sea surface and is not necessarily indicative of biological behavior. Differences in flight height are apparent between the small and big clusters (Figure  \ref{flight height stationarity}a). While tracks in the big cluster were concentrated at lower altitudes (82.3\% below 100 m), tracks in the small cluster were more frequently at higher altitudes and throughout the RSZ (37.9\% below 100 m). 

Flight height distributions varied with mean wind speed. Figure \ref{flight height stationarity}b,c shows flight height cumulative distributions conditional on mean wind speed. More animals were detected at higher altitudes during periods of faster wind speeds. For tracks in the big cluster, when wind speeds were below 2.3 m s\textsuperscript{-1}, 12\% of tracks were above 100 m compared to 32.4\% when wind speeds exceeded 13.7 m s\textsuperscript{-1}. This shift is smaller for the small cluster (54.6\% of animals flying above 100 m at wind speeds below 2.3 m s\textsuperscript{-1} and 69\% at wind speeds above 13.7 m s\textsuperscript{-1}). 

\begin{figure*}[t!]
    \centering
    \includegraphics[width=1\linewidth]{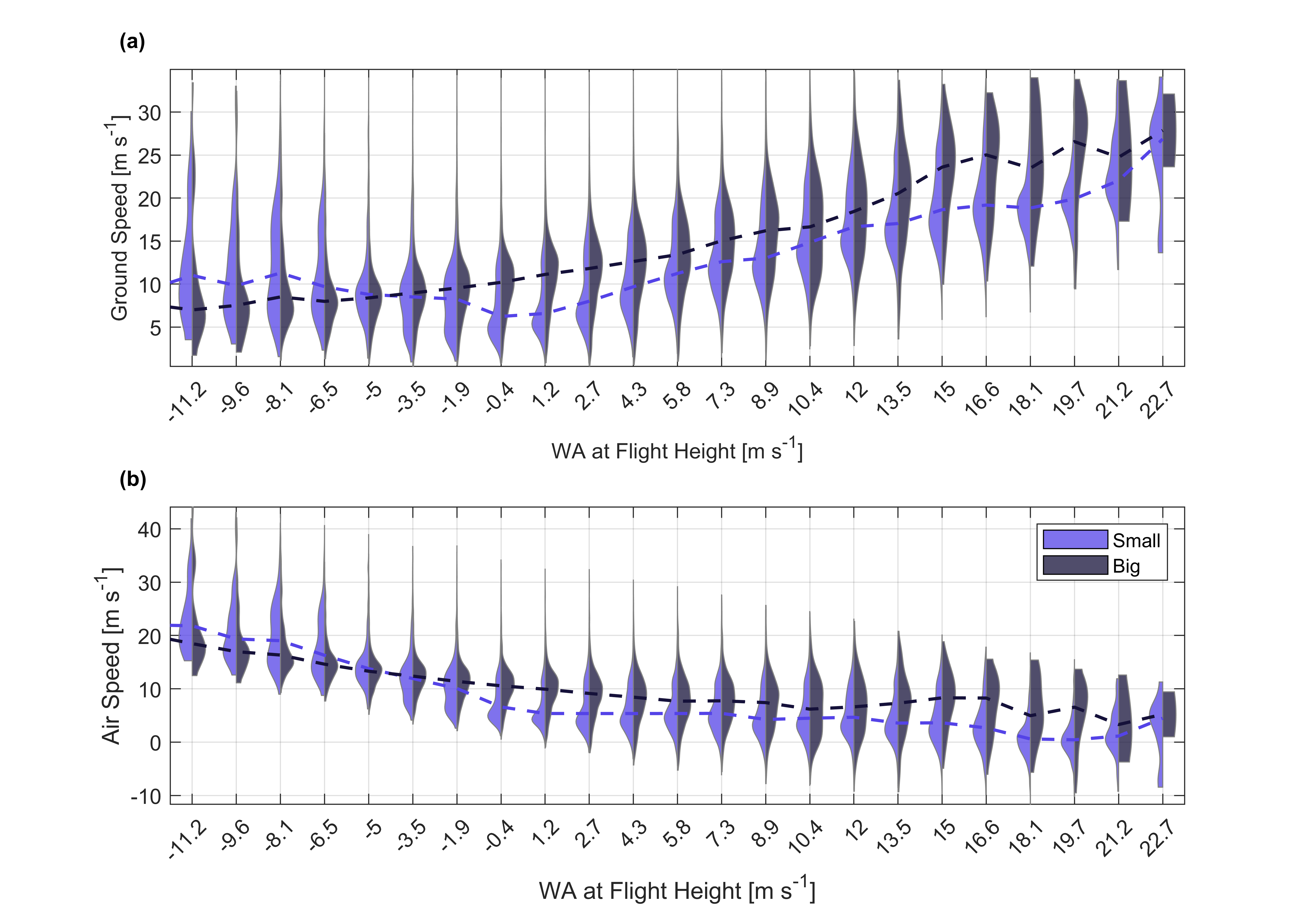}
    \caption{Violin plots of the relationship between (a) ground speed and (b) air speed with WA. Distributions are conditional on size cluster. Dashed lines are the median. }
    \label{speed}
\end{figure*}

Figure \ref{Wind assist}a,b shows the distributions of the height of maximum wind assistance for all tracks in each cluster.  Tracks in the small cluster more commonly had a height of maximum wind assistance above 100 m (88.2\%) than tracks in the big cluster (78.8\%). For both clusters, the proportion of tracks with a height of maximum wind assistance in the lowest altitude bin (40 to 80 m) corresponds primarily to tracks flying in headwinds and crosswinds. Note, wind measurements were not available below 40 m, so it is possible that the true height of maximum wind assistance for some of these tracks was at altitudes lower than 40 m. The mean wind profile shows that, on average, wind speed increased throughout the RSZ and decreased at higher altitudes. Figure \ref{Wind assist}c shows the difference between the height of maximum wind assistance and the measured flight height. Flight heights tended to be near the height of maximum wind assistance for both clusters, with 28\% of small cluster tracks and 22\% of big cluster tracks occurring within 50 m of their height of maximum wind assistance.

\subsection{Speed}
Violin plots of flight speed with wind assistance at flight height are presented in Figure \ref{speed}. For both clusters, ground speed is relatively consistent for tracks with negative wind assistance but increases with wind assistance for positive wind assistance (Figure \ref{speed}a). In contrast, air speed decreases as wind assistance becomes less negative and becomes relatively constant (at 5 m s\textsuperscript{-1} for the small cluster and 8 m s\textsuperscript{-1} for the big cluster) for positive values of wind assistance. Differences between the two clusters are observed. The clusters have similar median ground and air speeds for negative wind assistance values but diverge for most positive wind assistance values. Across all wind assistance conditions, animals in the big cluster tended to fly at slightly higher ground and air speeds than animals in the small cluster. Air speed can be considered as a proxy for energy input by an animal to flight, so higher airspeeds from the big cluster potentially signify higher energy input by larger animals resulting in faster ground speeds under the same wind assistance conditions.  

\section{Discussion}
Coupling a 3D bird and bat radar system with wind lidar provides unprecedented insights into flying animal behavior offshore. In this study, we show that wind speed and direction influence bird and bat behavior off the northeast coast of the United States. The high-resolution, co-located lidar measurements enabled identification of fine-scale relationships between flight behavior and wind speed and direction that would not be possible using modeled wind data (e.g., ERA5 reanalysis). Generally, animals were least likely to be present during high winds, and flight direction, height, and speed all varied with wind conditions. Our results are consistent with prior studies conducted in the North Sea. In line with \cite{Sjollema2014}, animal abundance at the site decreased at high wind speeds. Additionally, animals most frequently flew near the altitude with the highest wind assistance, a finding that is consistent with \cite{Bradaric2024}. We also observed changes in flight behavior with wind conditions, which is consistent with prior studies that have observed differences in flight style and behavior at different wind speeds \parencite{Ainley2015, gibb2018, kumagai2023,Skov2026}. Specifically, the proportion of animals flying with tailwinds, and at higher altitudes, increased with wind speed, and flight ground speeds increased with wind assistance despite air speed remaining mostly constant. In contrast to \cite{Manola2020} and \cite{Lagerveld2023}, we did not find alignment between wind and migration direction to be a strong predictor of animal presence, but did observe that animals predominantly flew with tailwinds. 

\subsection{Cluster Species Composition and Behavior}
Because it was not possible to determine the size or species of individual tracks in the radar data, we employed a clustering algorithm to divide data into two coarse size groups based on flight behavior. This resulted in a ``small'' cluster with lower values of radar reflectivity and a ``big'' cluster with higher values of radar reflectivity. We hypothesize that the small cluster is representative of smaller-bodied birds and bats and contains most migratory species, while the big cluster is representative of larger-bodied birds and contains most partial or short-distance migrants. Specifically, the four largest species expected to be present are non-migratory and are present year round, while the smallest species (wingspans less than 200 mm) are nearly all long-distance migrants (Figure \ref{species}). However, we expect that some large long-distance migratory species, such as the Great Skua (\textit{Stercorarius skua}), are included in the big cluster, and some non-migratory smaller species, like the Audobon's Shearwater (\textit{Puffinus iherminieri}), are included in the small cluster. Midsized resident species, such as Laughing Gulls (\textit{Leucophaeus atricilla}), may also be included in the small cluster. Distinct trends in animal behavior are observed that are consistent with the hypothesized cluster composition: mostly migrants in the small cluster and mostly residents in the big cluster. In particular, there were more short bursts in abundance in the small cluster (migratory pulses), and there was more consistent activity observed in the big cluster (Figure \ref{heatmaps}d). Additionally, animals in the small cluster had concentrated flight direction distributions that were aligned with migratory paths (Figure \ref{weekly wind and flight polars}) and flew at a variety of altitudes. In contrast, animals in the big cluster flew in a wide variety of directions but were concentrated at low altitudes (Figures \ref{conditional flight direction} and \ref{flight height stationarity}).  

Abundance modeling (Table \ref{modeling results}) revealed that the number of animals detected in the previous hour (animals prior) was the strongest predictor for both clusters, but accounted for a higher percentage of deviance explained for the small cluster (48.5\%) than for the big cluster (36.5\%). Conversely, the environmental predictors (wind speed, wind direction, wind turbulence intensity, sun azimuth, and sun elevation) accounted for a higher percentage of the deviance explained for the big cluster (7.32\%) than for the small cluster (4.56\%). These differences between clusters may be explained by the migratory behavior of animals in the small cluster---once a migration event begins (perhaps driven by biological behavior or other external factors not considered here), it is likely that abundance will be high in the proceeding hours until it tapers off despite potential changes in the environmental predictors. Future efforts could consider derived parameters, like rate of change, to capture the effect of a migration event ramping up versus ramping down. 

The effect of wind speed on abundance differed between the two clusters. In particular, wind speed had a positive effect on abundance out to higher wind speeds for the small cluster than the big cluster (positive influence below 20 m s\textsuperscript{-1} and 15 m s\textsuperscript{-1} for the small and big clusters, respectively, Figure \ref{PDP}d,f). Two possible and potentially coexisting explanations are (1) migratory species in the small cluster are compelled to fly at higher wind speeds while resident species in the big cluster can afford to be more picky about flight conditions, and (2) animals in the small cluster commonly fly with tailwinds and therefore could take advantage of higher wind assistance at higher wind speeds while those in the big cluster flying in headwinds would be increasingly penalized.  

In several cases, trends in the partial dependence plots (Figure \ref{PDP}) contradict trends in hourly abundance that are apparent in Figure \ref{abundance overview}a. For both size clusters, measured hourly abundance is lowest in non-daylight hours, but both models predict a positive association between abundance and sun azimuths corresponding to non-daylight hours. This apparent contradiction is potentially explained by the combined influence of sun azimuth and wind speed. Perhaps in the dark hours wind speeds are high enough to deter some flights, even though the time of day is advantageous for flight. The opposite may be true during morning daylight hours---favorable winds but less advantageous time of day. Then, as the winds begin to increase in the afternoon, higher wind speed and midday sun both repress flight. 



Track flight direction varied with both wind speed and direction. Similarly, \cite{Skov2026} showed wind speed and relative wind direction affected the angle of approach of Herring Gulls and Black-legged Kittwakes to offshore wind turbine rotors in the North Sea. In our work, both size clusters were more likely to fly with tailwinds when they were flying during periods with high wind speeds (Figure \ref{conditional flight direction}). This trend is also apparent in the polar plots of flight direction for each week of the study window (Figure \ref{weekly wind and flight polars}). Flight directions for tracks in the big cluster were more concentrated in the last two weeks of the study window when the highest wind speeds were measured (Figure \ref{heatmaps}), indicating that during these strong winds, animals had a stronger preference to fly with tailwinds. It is intuitive that animals would be more likely to select tailwinds at higher wind speeds; the energy required to fly in headwinds increases with wind speed. Even so, a substantial number of animals in the big cluster were observed flying with crosswinds or headwinds at the highest wind speeds (26.2\% at wind speeds above 13.7 m s\textsuperscript{-1}). 

Small animals more closely followed the wind and appeared to more readily adjust flight direction in response to the wind conditions than big animals. A couple trends support this: (1) smaller animals had a stronger propensity to fly with tailwinds than bigger animals -- especially at the highest wind speeds (Figure \ref{conditional flight direction}) where 94.5\% of the small cluster flew with tailwinds and (2) the diurnal variation of wind direction is apparent in the flight directions of tracks in the small cluster (Figure \ref{conditional flight direction hour of day}). These trends may be a consequence of wing and body sizes: Larger animals in the big cluster may be physically able to input more power into flight \parencite{Alerstam2007}, enabling travel with unfavorable wind directions. This is consistent with the higher air speeds of animals in the big cluster (Figure \ref{species}). This trend may also be related to less migratory behavior by animals in the big cluster; resident animals may not need to be as conservative with their energy as migratory animals.  

Differences in flight height distributions were apparent between the two clusters (Figure \ref{flight height stationarity}). Flight heights of animals in the small cluster were spread across altitudes below 500 m, while tracks in the big cluster were concentrated below 100 m. Wind assistance may partially explain these differences. Animals in both clusters most commonly flew near their height of maximum wind assistance. Under crosswinds and headwinds, the height of maximum wind assistance will typically be closer to the sea surface where wind speeds are lowest (Figure \ref{Wind assist}). Under tailwinds, the height of maximum wind assistance will typically be at higher altitudes with faster wind speeds. Animals in the big cluster more frequently flew with crosswinds or headwinds (50.9\%), so it follows that they flew lower. Animals in the small cluster more frequently flew with tailwinds (72.4\%) which implies they would fly higher to take advantage of faster wind speeds at higher altitudes. Further, animals may fly higher as wind speed increases (Figure \ref{flight height stationarity}) because the proportion of animals flying with tailwinds increases, and more animals can take advantage of faster, high-altitude winds. This is inline with the findings of \cite{kumagai2023} which showed that Black-tailed Gulls (\textit{Larus crassirostris}) flew higher under stronger tailwinds, \cite{Skov2026} which showed Herring Gulls and Black-legged Kittwake flew higher in strong tailwinds winds with low turbulence,  and \cite{TARROUX201699} which showed that Antarctic Petrels (\textit{Thalassoica antarctica}) flew lower under strong headwinds. Differences in flight height trends between clusters could also result from biological preference. For example, resident birds in the big cluster may choose to fly lower and in more directions because they are foraging while the migrating birds in the small cluster may fly higher for better visibility \parencite{TARROUX201699} and in a narrower direction band aligned with the migratory path. 

\subsection{Limitations of Radar Data}

The radar field of view did not extend below approximately 30 m. Therefore, the results and trends discussed in this work are descriptive of animal abundance and behavior within and above the RSZ, but do not capture the behavior of animals at lower altitudes. Without knowledge of how many animals were at lower altitudes, it is not possible to compute the percentage of animals present at the site that were within the RSZ. However, prior studies have indicated that there may be a significant number of birds and bats flying at altitudes below 30 m altitude \parencite{Brabant2019,Leemans2022,Schneider2024}. Additionally, it is likely that animals at higher altitudes were undercounted because radar detection capabilities roll off with range, and the maximum detection range will be shorter for animals with lower reflectivity. Specifically, the decreasing number of detected targets at flight heights above the RSZ for the small cluster may not be entirely representative of biological trends and could be the result of measurement bias. Figure \ref{heatmaps}a offers some insight into the relationship between reflectivity and detection range. The lowest reflectivity targets ($<-20$ dBsm) were not detected above an altitude of 300 m, while high reflectivity targets ($>5$ dBsm) were detected, albeit infrequently, at altitudes above 1000 m. The maximum altitude of detected targets increases approximately linearly with reflectivity between these two extremes. Lower reflectivity targets ($<-12$ dBsm) were frequently detected near their maximum altitude (i.e., pink region in the heat map), indicating that the upper limit of flight height may represent the upper limit of detection capabilities for these targets. Conversely, higher reflectivity targets ($> -12$ dBsm) were infrequently detected at higher altitudes (i.e., dark blue in the heat map), but there is a ``hot spot'' of higher reflectivity targets at lower altitudes. This indicates that limitations of radar detection range likely had a more significant influence on our analysis of targets with lower reflectivity (the small cluster). 

\subsection{Implications of This Study}

Not all collision risk models consider flight paths that are non-oblique to the RSZ, animal ground speed, or changes in animal behavior with wind conditions \parencite{Masden2016,Masden2021}. Our results support prior work indicating that these factors may be important considerations in collision risk modeling. \cite{Holstrom2011} discuss how collision risk is affected by the apparent frontal area of the turbine and flight path distance through the RSZ. Because the turbine will yaw such that the RSZ is perpendicular to the dominant wind direction, the apparent frontal area of the turbine will be highest for flight in headwinds or tailwinds and the flight path distance through the RSZ will be highest for flight in crosswinds. While most animals in this study flew with tailwinds, there were also substantial numbers of animals flying with crosswinds and headwinds, and this distribution changed with wind speed. \cite{Weiser2024} highlight how the proportion of animals within the RSZ, and therefore at the highest risk for collision, can vary with meteorological conditions. This is clear in our results where the proportion of animals in the RSZ generally decreased with wind speed (Figure \ref{flight height stationarity}); 83.5\% of tracks in the RSZ at wind speeds less than 2.3 m s\textsuperscript{-1} compared to 73.7\% when wind speed exceeded 13.7 m s\textsuperscript{-1}. Lastly, in support of \cite{Masden2021} who show that air speed estimates are not an accurate proxy for ground speed in collision risk models, our results show that ground speeds increase with wind assistance at flight height while air speed is relatively constant. In summary, collision risk models that consider flight direction, flight height, and ground speed as constants for a given species likely do not capture important trends that may impact collision risk assessments. 

In addition to changes in animal behavior, it is also necessary to consider how changes in turbine operation with wind speed may impact collision risk. \cite{marques2014} discuss how collision risk can increase with wind speed since, as a wind turbine spins faster, blade impact speed and strike probability increase while visibility of the blade likely decreases. Our results indicate the proportion of animals flying in the RSZ declines as wind speed increases, but that wind speed positively impacts abundance into high Region III wind speeds where turbine rotation rates are likely the fastest and animal flight speeds the quickest. These results paint a complicated collision probability picture. On one hand, collision risk could be elevated in Region III because of turbine operation and faster flight speeds that increase collision probability and the impact force. But, on the other hand, quicker transit times through the RSZ (due to faster flight speeds) and a lower proportion of animals in the RSZ could reduce collision probability in Region III. Further study is necessary to understand the aggregate impact of these parameters through all turbine operation regions. 


Finally, while our study is limited to a 5-week period in autumn 2024, and animal behavior may differ during other time periods or when offshore structures like wind turbines are present, the methodological approach presented lays the groundwork for future studies of the relationship between flying animal behavior and meteorological conditions. Our results indicate that combined analysis of radar and lidar data can reveal behavioral differences between animal size groups and offer new insights into how wind conditions influence animal presence, flight direction, flight height, and flight speed. Longer-term datasets of this kind could broaden the understanding of bird and bat behavior offshore and enable the development of predictive models for collision risk forecasting.

\section{Acknowledgments}
The authors wish to thank Jenny Wehof and Raghavendra Krishnamurthy at PNNL for their support of this work, as well as Jesse Lewis, Adam Kelly, and Jenny Davenport at DeTect, Inc for their assistance in interpreting the radar data. Additionally, we thank Kate Williams, Andrew Gilbert, Evan Adams, and Rebecca Stanley at the Biodiversity Research Institute for the information they provided on the species expected at the study location.

\section{Statements \& Declarations}

\subsection*{Funding}
This work was authored by staff at the Pacific Northwest National Laboratory operated for the U.S. Department of Energy (DOE) via Battelle (contract no. DE-AC05-76RL01830). This work was authored in part by NREL for the U.S. Department of Energy (DOE), operated under Contract No. DE-AC36-08GO28308. Funding provided by U.S. Department of Energy Office of Energy Efficiency and Renewable Energy Wind Energy Technologies Office. The views expressed in the article do not necessarily represent the views of the DOE or the U.S. Government. The U.S. Government retains and the publisher, by accepting the article for publication, acknowledges that the U.S. Government retains a nonexclusive, paid-up, irrevocable, worldwide license to publish or reproduce the published form of this work, or allow others to do so, for U.S. Government purposes.

\subsection*{Author Contributions}
All authors contributed to the study conception and design. Material preparation and data analysis was performed by Abigale Snortland and Emma Cotter. Bird and bat expertise were provided by Jeff Clerc and Cris Hein. The first draft of the manuscript was written by Abigale Snortland and Emma Cotter and all authors commented on previous versions of the manuscript. All authors read and approved the final manuscript.

\subsection*{Conflict of Interest}
The authors have no relevant financial or non-financial interests to disclose.

\subsection*{Data/code Availability}
The datasets generated during and/or analyzed during the current study will be made available in the Department of Energy's Wind Data Hub repository, https://wdh.energy.gov/ds/wfip3/barg.aviantracks.z01.c0

\subsection*{Ethics Approval}
Ethical approval was not applicable to this study as no animal test subjects were involved in this study.

\printbibliography

\end{document}